\documentstyle{article}

\newcommand {\tsub}[1]{_{\mbox{\protect\scriptsize #1}}}
  \newcommand 
{\Ref}[1]{(\ref{#1})}

\newcommand {\diracm}[1]{\left|#1\right\rangle}
 
 \newcommand {\Stretch}[1]{
\renewcommand{\baselinestretch}{#1 }\large\normalsize}
 
\newcommand {\bm}[1]{{\mbox{\boldmath $#1$}}}

\newcommand {\citex}[1]{\protect\cite{#1}} 

\def\beq{\begin{equation}}
\def\eeq{\end{equation}}
\def\bea{\begin{eqnarray}}
\def\eea{\end{eqnarray}}

\def\ag{A_g^+}
\def\bu{B_{1u}^-}

\def\s0{\phi_0}

\def\n2{\nabla^2}

\begin{document}

\Stretch{3}

\setcounter{page}{1}
\baselineskip=25pt \parskip=0pt plus2pt
\begin{center}
{\large \bf Density Matrix Renormalisation Group Calculations of the
Low-lying Excitations and Nonlinear Optical Properties of 
poly({\em para}-phenylene)}

\Stretch{1.5}

{William Barford$^1$, Robert J.\ Bursill$^2$ 
and Mikhail Yu.\ Lavrentiev$^{3*}$ \\ }
$^{1,3}$Department of Physics, The University of Sheffield, 
Sheffield, S3 7RH, UK \\
$^2$School of Physics, University of New South Wales, Sydney, 
NSW 2052, Australia\\

{\small
Email: $^1$W.Barford@sheffield.ac.uk,
$^2$ph1rb@newt.phys.unsw.edu.au,
$^3$M.Lavrentiev@sheffield.ac.uk}


\end{center}
\Stretch{2}

\begin{abstract}
The two state molecular orbital (2-MO) model of the phenyl based
semiconductors
is used to calculate
the low-lying spectra of the $A^+_g$ and $B^-_{1u}$ states of
poly({\em para}-phenylene) (PPP).
The model parameters are
 determined by 
fitting its predictions to exact Pariser-Parr-Pople model calculations 
of benzene and biphenyl, and it is solved using the density matrix
 renormalisation group method.
It is shown that there exists a band of $^1\bu$ (`s'-wave)
excitons below the
 band states. 
In the long chain limit the lowest exciton is situated $3.3$ eV above the
ground state, consistent with experimental data. 
The calculated
particle-hole separation
of these excitons indicates that they are tightly bound, 
extending over only a few repeat units.
The lowest
band state is found to be a covalent 
$2^1\ag$ state, whose energy almost
coincides  with the charge gap $E_G$. 
Lying just above the $2^1\ag$ state is a band $^1\bu$ 
state (the $n^1\bu$ state).
The  particle-hole separation of the
band states scales linearly with oligomer size. 
The binding energy of the $1^1\bu$ exciton is 
determined rigorously as $0.74$ eV.

The dipole matrix elements and oscillator strengths 
for the transitions between the lowest $^1\ag$ and $^1\bu$ states 
are calculated and 
the NLO properties of PPP, such as electroabsorption (EA) and 
third harmonic generation, are investigated. A comparison of the 
EA spectrum with the experimental data
 show that the main features of the experimental 
spectrum are well described in the 2-MO Hamiltonian. 

Only five states account for most of the calculated EA. 
These are the $1^1\ag$, $1^1\bu$, $2^1\ag$, $n^1\bu$ and another band 
$^1\ag$ state, the $k^1\ag$,
thus confirming the essential states model.  An analysis of the particle
excitation weight of these states indicates that they are predominately
single particle in character.

\end{abstract}



\pagebreak

\section{Introduction}

Since the first light-emitting device based on 
poly($para$-phenylenevinylene) (PPV) was reported \cite{nature90}, 
the non-linear optical (NLO) properties of conjugated polymers have been 
extensively investigated. Amongst the numerous systems 
studied, poly($para$-phenylene) (PPP), being a linear chain of phenyl 
rings, possesses one of the simplest structures. However, its electronic 
structure and the nature of the blue light emission \cite{grem92} are 
still controversial. First-principles local-density approximation studies 
by Ambrosch-Draxl $et$ $al.$
\cite{ambroschdraxl95} suggest that the optical properties of PPP can be 
explained by a purely band picture, with intragap non-linear excitations 
suppressed by three-dimensional effects. However, recent experimental 
results on the electroabsorption (EA) and photoinduced absorption (PA) 
in substituted PPP by Lane $et$ $al.$ \cite{lane97} are explained by 
the presence of non-linear excitations, such as singlet and triplet 
excitons, and charged polarons. 

The aim of this paper is to clarify the r\^ole and importance of the 
low-lying
non-linear excitations in PPP by calculating its electronic 
structure and NLO properties in a realistic Hamiltonian. 
The EA spectrum compares favourably with recent experiments. 
We identify the key states which participate in the NLO processes. 
Moreover, by calculating the particle-hole separation of these states, 
we identify the band gap as the threshold state whose particle-hole 
separation increases linearly with oligomer size. This enables a 
rigorous determination of the band gap to be made.

Recently, a two state molecular orbital model was
introduced \cite {barford97a} to describe the $B_{1u}$ and $A_g$ states of the
phenyl based semiconductors.  In the current paper we introduce a more
thorough parameterisation of this model by fitting to improved 
exact Pariser-Parr-Pople model calculations of the molecular
building blocks (i.e. benzene and biphenyl) \cite{bb3}. This
model is then solved for oligomers of arbitrary length without
further parameterisation.

As well as our earlier work, which was the first to use the DMRG
method for the phenyl based semiconductors
 \cite{barford97a, barford97b},  there have been a number of other
theoretical calculations on PPP. Br\'{e}das has used the VEH 
pseudopotential technique \cite{bredas2}, Champagne et al.\ have
 performed Hartree-Fock calculations \cite{champagne96}, 
and Ambrosch-Draxl et al.\ have performed 
density functional calculations using LAPW and pseudopotentials
 \cite{ambroschdraxl95}. Rice 
et al.\ \cite{rice} have developed a phenomenological, microscopic
 model based on the 
molecular excitations of benzene. The absorption bands are calculated 
using an 
approximate Kubo formalism.
In a series of papers, Shimoi and Abe have considered
the optical and electro absorption of PPV 
in a re-parameterised P-P-P model using single-excitation
configuration interaction (S-CI) \cite{shimoi96, shimoi97}.
This method was used also by Harigaya {\em et al.}
 for studying optical
absorption spectra and exciton properties in PPP, PPV,
and related polymers \cite{harigaya97}.

The structure of a PPP chain is shown schematically in Fig.\ 1. 
It possesses $D_{2h}$ symmetry.
The electronic states can be classified according to their spatial, 
spin and particle-hole symmetries.
In this paper we only consider states which are symmetric under reflection
in the $x-z$ plane, and either symmetric or anti-symmetric under reflection
in the $x-y$ plane. These will correspond to the low energy excitations.
The ground state belongs to the $^1\ag$
(spin singlet, space- and particle-hole-symmetric) symmetry sector. 
Low-lying one-photon excitations
occur between the $^1\ag$ and the $^1\bu$
(spin singlet, odd space and 
particle-hole symmetry) symmetry sectors. These excitations are
polarised along the long ($z$) axis.
We note that these states are {\em electronically decoupled} from the 
higher lying $B_{2u}$ (symmetric under $x-y$ reflection and anti-symmetric
under $x-z$ reflection)
and $B_{3g}$ (anti-symmetric under $x-y$ and $x-z$ reflection) states.

Non-linear processes in polymers with inversion symmetry are determined 
by the third-order susceptibility 
$\chi^{(3)}(-\omega_1-\omega_2-\omega_3;\omega_1,\omega_2,\omega_3)$, 
which can be calculated
as a sum over intermediate states (see, e.g., \cite{orr71}). It was 
suggested several years ago that most of the NLO properties can be described 
by an ``essential states'' model \cite{dixit91, guo93, mazumdar94, yaron96}. 
The concept of the essential states model implies that there is a 
restricted set of states responsible for most of the NLO properties of 
the system. Usually, these are the ground state ($1\ag$), the lowest 
excited (excitonic) $^1\bu$ state, 
the $m^1\ag$ state, which is the $^1\ag$ state most strongly related 
to the $^1\bu$ states through one-photon excitations,
and the band threshold $n^1\bu$ state.
We investigate the validity of this approach for PPP. 

Until recently, numerical investigations of one-dimensional systems 
with strong electron-electron interactions were limited to exact
diagonalisations of short chains, to approximate configuration 
interaction calculations
or to the use of density functional theory. A key aspect of this work 
is that we perform essentially exact calculations
on our model using the density matrix renormalisation group 
(DMRG) method. We check the numerics by comparing DMRG results with 
exact results in the non-interacting limit, by comparing DMRG and exact 
diagonalisation calculations for a 7 unit oligomer and by monitoring 
the convergence of results with the parameter $m$ (which controls 
the amount of Hilbert space truncation and hence the DMRG error) 
for longer systems in the interacting case.

The plan of this paper is as follows. In the next section the 
molecular orbital model will be reviewed and parameterised. In \S 3 
the energy and root-mean-square size of the low-lying states are 
calculated as a function of oligomer size, thus identifying the band 
threshold.  The single particle excitation weight of the 
wavefunctions is also investigated.
 In \S 4 the dipole moments between states are calculated, 
thus identifying the key states in the NLO processes. The linear 
susceptibility is also calculated. In \S 5 we turn to the calculation 
of the EA and THG spectra using the sum-over-states method. Finally, 
we conclude in \S 6.

\section{The Molecular Orbital Model and Hamiltonian} 

The starting point for the molecular orbital approach used in this paper is
the well-known Pariser-Parr-Pople or extended Hubbard Hamiltonian: \beq
H = - \sum_{<ij>\,\sigma} t_{ij} \left[ c_{i\sigma}^{\dagger} c_{j\sigma} +
\rm{ h.c.} \right] + U \sum_{i} \left( n_{i\uparrow} - \frac{1}{2} \right) 
\left(
n_{i\downarrow} - \frac{1}{2} \right) + \frac{1}{2} 
\sum_{i\neq j} V_{ij} (n_i - 1)(n_j - 1), \label{ppp_hamiltonian}
\eeq
Here, $c_{i\sigma}^{\dagger}$ and $c_{i\sigma}$ are creation and destruction
operators, respectively, for a $\pi$ electron with spin $\sigma$ 
on carbon site $i$, $n_{i\sigma} = c_{i\sigma}^{\dagger}c_{i\sigma}$ is the
number operator, $n_{i} = n_{i\uparrow} + n_{i\downarrow}$ and $t_{ij}$ is 
the
transfer integral. $U$ and $V_{ij}$ are Coulomb repulsion
 parameters for electrons occupying one site and sites $i$ and $j$, 
respectively. 

The Ohno parameterisation for the Coulomb interaction is,
\begin{equation}
V_{ij} = \frac{U}{ (1 + \alpha r_{ij}^2)^{1/2} },
\end{equation}
where $\alpha=(U/14.397)^2$, thus ensuring that $V_{ij} \rightarrow 
e^2/(4\pi\epsilon_0 
r_{ij})$ as $r_{ij} \rightarrow \infty$, and $r_{ij}$ is the
 inter-atomic distance in \AA. The optimal parameterisation for PPP,
which was 
derived in \cite{bb3}, is $U=10.06$ eV,the phenyl bond transfer integral, 
$t\tsub{p}=2.539$ eV and the single bond transfer integral,
$t\tsub{s}=2.22$ eV.

\subsection{The Model}

The six atomic $\pi$-orbitals of a phenyl ring $c_{i\sigma}^{\dagger},
 i=1,...,6$
may be transformed to six molecular orbitals (MOs), 
$a_{\alpha\sigma}^{\dagger}$.
This approach for phenyl based semiconductors was used earlier by 
Soos {\em et al.} \cite{soos92}. Recently, 
Chandross {\em et al.} \cite{chandross97} employed a similar 
approach in their work on the
characterization of excited states in conjugated polymers. 
It was suggested in \cite{barford97a} that the low lying 
$A_g$ and $B_{1u}$ excitations 
can be described by only two of the resulting MOs, namely, 
the bonding $e_{1g}$ HOMO and
$e_{2u}$ LUMO states.
This is the so-called two state molecular orbital (2-MO) model.
The other HOMO and LUMO states are non-bonding, because the wave function 
amplitude on the bridging carbon atoms is zero.
Transitions between these states lead to high lying
localised $B_{1u}$ excitations, while transitions which mix the bonding and
non-bonding orbitals lead to excitations with $B_{2u}$ and $B_{3g}$
symmetry \cite{bursill97}. We will not be concerned with
these latter
 states in this paper. The (occupied) $a_{2u}$ 
and (empty) $b_{2g}$ states are situated far away in energy from the 
HOMO and LUMO, and are assumed to play only a minor role in the
low-energy excitations.
Below, the bonding
HOMO orbital is denoted by $\diracm{1}$ and the bonding LUMO orbital 
is denoted by $\diracm{2}$.

With two MOs left, and neglecting the
three and four centre two-electron integrals and Coulomb interactions 
beyond nearest neighbour phenylene units,
the transformation from the atomic to the molecular orbital basis results 
in the following Hamiltonian:
\begin{eqnarray}
H & & = -\sum_{i\,\alpha\,\beta\,\sigma} t_{\alpha\beta}
 \left[ a_{i\alpha\sigma}^{\dagger} a_{i+1\beta\sigma}
 + \rm{ h.c.}\right] + \sum_{i\,\alpha}
\epsilon_{\alpha}(n_{i\alpha}-1) + U \sum_{i\,\alpha}
 \left(n_{i\alpha\uparrow}-
\frac{1}{2}\right) \left(n_{i\alpha\downarrow}-
 \frac{1}{2}\right) \nonumber \\
& &
+\;\frac{U}{2} \sum_{i\,\alpha\neq\beta} (n_{i\alpha}-1)
 (n_{i\beta}-1) + V \sum_{i \,\alpha\,\beta} (n_{i\alpha}-1)
 (n_{i+1\beta}-1) \nonumber \\
& &
-\;X \sum_{i\,\alpha\neq\beta}
\left[ \bm{S}_{i\alpha}.\bm{S}_{i\beta} +\frac{1}{4}
\left(n_{i\alpha}-1\right)
\left(n_{i\beta}-1\right) \right]
\nonumber\\
& &
+\;\frac{P}{2} \sum_{i\,\alpha\neq\beta\,\sigma}
 a_{i\alpha\sigma}^{\dagger}
a_{i\alpha\bar{\sigma}}^{\dagger} a_{i\beta\bar{\sigma}} 
a_{i\beta\sigma},
\label{model},
\end{eqnarray}
where
$\bm{S}_{i\alpha}=\sum_{\rho \rho^\prime}
a_{i\alpha\rho}^{\dagger}
 \bm{\sigma}_{\rho\rho^\prime}a_{i\alpha\rho^\prime}$ 
and $\bm{\sigma}$ are the Pauli 
spin matrices. 

The key interactions in this model are:
 the HOMO-LUMO gap 
($\Delta = \epsilon_2-\epsilon_1$), 
direct onsite ($U$) and nearest neighbour ($V$) MO Coulomb repulsion,
spin-exchange ($X$)
and pair hop ($P$)
 between MOs on the same repeat unit, and hopping ($t$) between neighbouring
repeat units.  To understand the essential physics of this model, consider
the limit $t=0$. In this limit there are localised intra-phenyl
particle-hole triplet and singlet excitations at $\sqrt{\Delta^2 + P^2}
-X$ and $\sqrt{\Delta^2 + P^2}+X$, respectively. The MO Coulomb 
repulsion results in a potential well to the unbinding of these excitations.
It costs 
an energy $U-V-X$ to seperate the particle-hole pair by one repeat unit
and $U-X$ to separate them by two or more repeat units.  Finally, the
hybridisation, $t$, leads to the delocalisation and ultimately unbinding
of the particle-hole pair.

A straightforward derivation of the new Hamiltonian parameters from
the atomic Hamiltonian (1) gives results for the excitation energies which
deviate from exact Pariser-Parr-Pople model calculations of
benzene and biphenyl, as well as to over-estimating the optical gap
by approximately $1$ eV in long 
oligophenylenes
\cite{bursill97}.
We therefore take the view that eqn. (3) contains the essential physics
to model the low-lying excitations,
but that these interactions are renormalised from their bare 
Pariser-Parr-Pople values.
We parameterise 
the two state model by fitting its predictions
to the exact
Pariser-Parr-Pople model calculations of benzene and biphenyl \cite{bb3}. 
The two state model is exact in the limit that the interactions vanish,
as in that case it describes particle-hole excitations from the
valence to the conduction band.  In the other extreme of the hybridisation
vanishing, it correctly models localised intra-phenyl triplet and
singlet excitations. In the intermediate regime the validity of the
approach is determined
by the test with experiment.  We will  show that the predictions
are in good agreement with the experimental data, but in addition
substantial physical insight is achieved as to the nature of the
excited states.

\subsection{Parameterising the Model}

Since the 2-MO model is applicable to states of $A_g$ and $B_{1u}$
symmetry, it is parameterised by fits to states of the same symmetry
in benzene and biphenyl.
The interactions are parameterised in the following way: First, when the 
nearest neighbour hybridisation is switched off, the model should
predict localised (phenylene) triplet and singlet excitons.
The full Pariser-Parr-Pople calculation of benzene
predicts a pair of excitations which are anti-symmetric under
$x-y$ reflection (the $1E_{1u}(z)$ and $1B_{1u}$ states) and a pair
which are anti-symmetric under $x-z$ reflection (the $1E_{1u}(y)$
and $1B_{2u}$ states) in both the singlet and triplet channels \cite{bb3}.
However, as was explained
in \cite{bursill97},
the 2-MO model predicts that both 
the pair of $1E_{1u}(z)$ and $1B_{1u}$ states
and the pair of $1E_{1u}(y)$
and $1B_{2u}$ states are
 degenerate.
In the 2-MO model the 
energies of the triplet and singlet are $\sqrt{\Delta^2 + P^2}-X$ and 
$\sqrt{\Delta^2 + P^2}+X$, respectively. $\Delta$ and $X$ (since $P=X$) are
determined by fitting these values to the {\it average} values of the 
$1E_{1u}(z)$ and $1B_{1u}$
triplet and singlet excitons obtained from the 
full Pariser-Parr-Pople calculation on benzene, which are $4.45$ eV and $6.23$
eV, respectively \cite{bb3}.
 This gives $\Delta=5.26$ eV and $X=0.89$ eV. Next, when
the hybridisation is switched on the excitons delocalise and interact. The key
low-lying states are the long axis-polarised triplet ($1^3B^+_{1u}$) and 
singlet ($1^1B^-_{1u}$) states, and the lowest even, {\em covalent} 
excitation, the $2^1A^+_g$
\cite{biphenylnote}. Once $\Delta$ and $X$ have been fixed, their
energies are determined by $t$, $U$ and $V$ (we 
assume $|t_{\alpha\beta}| \equiv t$ for all orbitals $\alpha$, $\beta$). We
use the exact biphenyl calculations to fit these excitations. To simplify
the fitting of these parameters we assume that $V=U/2$, and adjust
$t$ and $U$ so that we have an exact fit to the full 
Pariser-Parr-Pople calculation of
the biphenyl $1^1B^-_{1u}$ state and a {\it minimum} relative error for the
$1^3B^+_{1u}$ and $3^1A^+_g$ states. This gives $U=3.67$ eV, 
$V=1.835$ eV, $t=0.895$ eV, and a relative error for the $1^3B^+_{1u}$ and
$3^1A^+_g$ states of $-2.2$\%. We note that since the band width is 
approximately equal to
$U$, these parameters are in the intermediate coupling regime. 
These parameters differ from those of \cite{barford97a}, as they are
based on a more rigorous fit to benzene and biphenyl.  For ease of
reference they are listed in table~\ref{parameters}.

\subsection{Solving the Model}

\subsubsection*{(i) Density matrix renormalisation group solution} 

The system \Ref{model} is a one dimensional quantum lattice model with 
16 states per repeat unit. For small lattice sizes, $L$, it is possible 
to calculate eigenvalues and eigenstates using exact diagonalisation.
However, the largest system which can comfortably be reached by this 
method is the sexamer ($L=6$ phenylene units). In order to study larger 
systems, we turn to the DMRG method \cite{white}. The DMRG is a powerful, 
robust, portable and highly accurate truncated basis scheme for the 
solution of low dimensional quantum lattice systems, and is especially 
well suited to the solution of open linear chains such as \Ref{model}. 
We have performed calculations of up to 10 low energy eigenvalues and 
eigenvectors in the $^1\ag$ and $^1\bu$ symmetry sectors, as well as 
dipole matrix elements, oscillator strengths and correlation functions 
for systems of up to 37 repeat units with sufficient accuracy to make 
comparisons with experiment. 

The DMRG is discussed at length in \cite{white} and reviewed in 
\cite{gehring} so we restrict our discussion here to features relevant 
to our implementation of the method for \Ref{model}. The key features 
are the form of the {\em system}, {\em environment} and {\em super} blocks, 
the number $m$ of states retained per block, and the good quantum numbers 
used to diagonalise the superblock Hamiltonian and the density matrix. We 
implement the DMRG for \Ref{model} using the {\em infinite lattice algorithm} 
\cite{white}. That is, the system and environment blocks are reflections 
of one another, and are increased by one repeat unit at a time, the initial 
blocks consisting of a single phenylene unit. The first two superblocks are 
schematically depicted in Fig.\ 2. They are comprised of the system and 
environment blocks abridged by a phenylene repeat unit.

The total charge $\hat{N} = \sum_{i\,\alpha}n_{i\alpha}$ and the 
total $z$ spin
$\hat{S}^{z}\tsub{T} = \frac{1}{2} \sum_{i\,\alpha} \left(n_{i\alpha\uparrow}
 - n_{i\alpha\downarrow}\right)$ 
are used as good quantum numbers in diagonalising the superblock hamiltonian 
and the system and environment block hamiltonians and density matrices. 
In addition, the spatial inversion 
($\hat{C}_{2}$: $a_{i\alpha\sigma} \mapsto a_{L-i+1\,\alpha\sigma}$),
particle-hole
($\hat{J}$: $a_{i1\sigma}^{\dagger}
\mapsto {\rm sgn}(\sigma)a_{i2\bar{\sigma}}$, $a_{i2\sigma}^{\dagger}
\mapsto {\rm sgn}(\sigma)a_{i1\bar{\sigma}}$) %
and spin flip
($\hat{P}$:
$a_{i\alpha\sigma} \mapsto a_{i\alpha{\bar\sigma}}$) %
symmetry operators can be constructed for the superblock. 
Their corresponding projection operators can be applied to random 
superblock states in order to construct trial states of definite 
symmetries which can be fed into the sparse matrix diagonalisation 
routine used in the diagonalisation of the superblock hamiltonian. 
The resulting target states retain these symmetries as long as the 
iterated trial state is periodically resymmetrised, for example, 
every 30 or so matrix multiplications.
This procedure is numerically stable because the density matrix 
eigenstates are eigenstates of the block symmetry operators and hence 
the obtained superblock energy eigenstates are {\em exact} 
(to within machine precision) eigenstates of superblock symmetry 
operators at each iteration. This procedure is checked for each 
calculated state by determining the expectation values of the 
symmetry operators, which are found to equal $\pm 1$ to around 
12 decimal places.

\subsection*{(ii) Accuracy tests}

We verify the validity of the DMRG solution by checking that the results 
obtained for the trimer and the pentamer agree with exact diagonalisation 
results. Basis truncation occurs for larger chains, the first being 
the septamer. This is the largest system that we can treat by exact 
diagonalisation (the 
dimension of the
Hilbert space is 11 778 624).
In Table \ref{comparison} we compare the exact results 
with two DMRG calculations (with 65 000 and 130 000 states retained) 
for the energies and excitons sizes of the $1^1\bu$ and $2^1\ag$ states 
of this system. It is clearly seen that, despite the fact that the DMRG 
calculation uses only a fraction of the total number of states 
(0.55\% and 1.10\%, respectively), in both cases the results are very 
close to the exact results, both for the energies and for the exciton sizes.

In order to check the convergence for longer systems we first examinine 
the non-interacting ($U = V = X = P = 0$) case which can easily be 
diagonalised exactly for any chain length. In the DMRG calculations 
we retain $m = 230$ states per block. Exact and DMRG results are given 
in Table~\ref{dmrgtest1} for the ground and first excited state energies 
for a number of lattice sizes. We see that the DMRG resolves gaps between 
these states well and truly above the accuracy required in order to make 
comparisons with experiments, that is, a few hundreds of an eV. The accuracy 
is expected to be even better in the interacting case where states are 
more localised and gaps are widened \cite{white}. That is, the accuracy 
should increase monotonically as $U$ is increased, up to the atomic 
limit $t_{\alpha\beta}=0$ where exact results are trivially recovered 
by any real space renormalisation group procedure. In Table 
\ref{dmrgtest2} we monitor the convergence of the optical gap in the 
interacting case with the truncation parameter $m$. We again see that 
the results have converged within errors which are negligible in so far 
as comparisons with experiment are concerned. Finally, DMRG calculations 
of the exciton correlation functions of the $1^1\bu$ and $2^1\ag$ states 
are checked for systems of up to 21 repeat units by doubling the size of 
the superblock Hilbert space. The change in the exciton size is found 
to be less than 1\% for $1^1\bu$ and 3\% for $2^1\ag$ for all the 
systems studied.

\section{The Low Energy Spectra and Exciton Correlation Functions} 

Table 4 shows the energies of the lowest triplet and singlet $B_{1u}$ 
excitons
as well as the lowest {\em covalent} even ($^1\ag$) excited state. 
The agreement between
the predicted results from the 2-MO model and experiment are good
for both oligomers and the polymeric limit, 
confirming the validity of our parameterisation.
The calculated energies of the lowest excited $^1\ag$ and $^1\bu$ 
states as a function of
oligomer size $N$ are given in Fig.\ 3. Also, the charge gap defined as 
$E_G = E(2N+1)+E(2N-1)-2E(2N)$,
is plotted. Here, $E(2N)$ is the ground state energy of a system with 
$2N$ electrons. The lowest excited state has $^1\bu$ symmetry, its energy 
always being
lower than the charge gap.
As the chain length is increased, the number of $^1\bu$ states below 
the charge gap also increases, creating an excitonic
band in the limit $N \rightarrow \infty$. (The $^1\bu$ states are inter-leaved
with $^1\ag$ states.) 
The energy of the
lowest excited $^1\ag$ state, $2^1\ag$, almost 
coincides with the charge gap $E_G$,
thus strongly implying that
$2^1\ag$ is a band threshold state. The energy of the lowest $^1\bu$ 
state above
the $2^1\ag$, denoted hereafter as $n^1\bu$, also tends to 
$E_G$ as $N \rightarrow \infty$.
A polynomial fit of the energies as a function of inverse
oligomer length indicates that in the limit 
$N \rightarrow \infty$, the energies of the $2^1\ag$ and the
$n^1\bu$ states, 
and $E_G$ all tend to a value of 4.04 eV. The energy of
the $1^1\bu$ state 
approaches
3.30 eV.
Thus, the energy results are good evidence that the lowest 
excited $^1\ag$ state is a threshold state dividing the spectrum 
into bound excitonic states below and unbound band-like states 
above it.

That the $2^1\ag$ state is a band state is confirmed by an examination
of the exciton spatial correlation function
defined as:
\begin{eqnarray}
C_{ij}(\left|n\right\rangle) = \left\langle{n}\right|{S_{ij}^{\dagger}}
\left|{1^1A_g^+}\right\rangle,
\label{correlation}
\end{eqnarray}
where
\begin{equation}
S_{ij}^{\dagger} = \frac{1}{\sqrt{2}} (a_{i2\uparrow}^{\dagger} 
a_{j1\uparrow} + a_{i2\downarrow}^{\dagger} a_{j1\downarrow})
\label{exciton}
\end{equation}
is a singlet exciton creation operator, which removes a particle from the 
orbital $\left|{1}\right\rangle$ 
on site $j$ and places it onto the orbital $\left|{2}\right\rangle$
on site $i$.  Thus,
\beq
W_1=\sum_{ij}C_{ij}^2
\label{weight}
\eeq
gives the weight of single particle excitations in the state $\diracm{n}$
and
\beq
P_{ij}= C_{ij}^2/W_1
\eeq
is the distribution function for the particle-hole separation.
Note that, as discussed in Appendix A,
$C_{ij}=+C_{ji}$ for states which are negative under the particle-hole 
transformation ({\em i.e.} `s'-wave excitons), while $C_{ij}=-C_{ji}$ 
for states which are positive under the particle-hole transformation 
({\em i.e.} `p'-wave excitons).

Using (7), we
calculate the spatial extent of a given state, or the particle-hole 
separation, using the formula:
\beq
l^2(\diracm{n}) = \langle(i-j)^2\rangle = \sum_{ij}(i-j)^2P_{ij}.
\label{rms}
\eeq
(A similar approach to calculating the average particle-hole separation was
used by D.\ Yaron and R.\ Silbey in their study of polyacetylene
 \cite{yaron92}.)
The results are given in Fig.\ 4 as a function of oligomer size $N$. 
The $^1\bu$ states belonging to the excitonic band below the charge 
gap are those with the smallest electron-hole distance. With increasing 
oligomer size, the particle-hole distance in the $1^1\bu$, as well as 
in other states in the excitonic band, approaches
a limit of approximately 
one repeat unit. This indicates that these are strongly bound `s'-wave
excitons. 
Conversely, the spatial extent of the 
$2^1\ag$ and $n^1\bu$ states are
proportional to the oligomer size, and scales in the same way 
with $N$ as do
the lowest excited (unbound)
states (namely, the $1\bu$ and $2\ag$ states)
of the {\em non-interacting} model.

We believe that all these results provide strong evidence that 
there exists an $^1\bu$ exciton in PPP, and that the lowest band state 
has the $^1\ag$ symmetry. The exciton binding energy for long 
oligomers approaches $E_G - E_{1^1\bu} = 0.74$ eV.

Finally, in Table 5 we show the single particle weight, $W_1$ eqn (6), for the
essential states (as defined in the next section)
 for a 15 site oligomer.  Evidently, these states
are predominately single particle in character.

\section{Oscillator Strengths and the Linear Susceptibility} 

As a next step towards the
calculation of the NLO properties of PPP, we compute the oscillator 
strengths of transitions between the essential states. These are, 
besides the ground state, the $1^1\bu$, the $2^1\ag$, the $n^1\bu$ 
and also the $^1\ag$ state situated above 
the $n^1\bu$,
which we denote as $k^1\ag$. This state has already been invoked 
for explaining the EA spectra of a number of luminescent and 
non-luminescent polymers by Liess $et$ $al.$ \cite{liess97}. 
In Fig.\ 5 we show schematically
the most important states for the non-linear optical properties.
The results for the corresponding oscillator strengths 
are given in Table 6. 

Using the results for the oscillator strengths,
 the first-order optical absorption is computed.
As expected,  the dominant peak belongs to the lowest
allowed transition, $1^1\ag \rightarrow 1^1\bu$, while 
the band threshold state 
($n^1\bu$) is represented by a weak feature at $4.3$ eV, as shown in Fig.\ 6
for a 15 site oligomer.  The higher lying optical transitions to states
of $B_{2u}$ symmetry are not indicated in this figure, as we do not
model these states in our theory.  The high lying localised
$1^1\ag \rightarrow ^1\bu$
transition is also not explicitly indicated, but as
discussed in \S 2, our theory predicts this to be at 6.23 eV.

\section{Third Order Non-Linear Susceptibilities} 

The NLO properties of PPP, such as third harmonic generation (THG) 
and electroabsorption (EA), can be related to the third-order 
macroscopic susceptibility $\chi^{(3)}$. 
The EA signal is related to the imaginary part of 
$\chi^{(3)}$ \cite{liess97}: 
\beq
-\Delta T/T = \frac{4 \pi \omega F^2 d}{c n} Im \chi^{(3)}(-\omega;
\omega,0,0),
\eeq
while the THG spectrum is related to
$\chi^{(3)}(-3\omega;\omega,\omega,\omega)$.
The third-order 
susceptibility $\chi^{(3)}$, in turn, results from the third-order 
microscopic hyperpolarizability $\gamma_{xxxx}$:
\beq
\chi^{(3)}_{xxxx}(-\omega_{\sigma};\omega_1,\omega_2,\omega_3) =
 \frac{1}{5} f^{\omega_\sigma}_{xx} f^{\omega_1}_{xx}
 f^{\omega_2}_{xx} f^{\omega_3}_{xx} \gamma_{xxxx},
\eeq
where $\omega_{\sigma} = \omega_1 + \omega_2 + \omega_3$ and the factor 
$\frac{1}{5}$ results from the orientational averaging \cite{neher}. 
The calculation 
of $\gamma_{xxxx}$ can be performed using the sum-over-states method 
(see, e.g., \cite{orr71}): 
\bea
\gamma_{xxxx}(-\omega_\sigma;\omega_1,\omega_2,\omega_3) =  
K(-\omega_\sigma;\omega_1,\omega_2,\omega_3) (-\hbar)^{-3} \nonumber \\
I_{1,2,3} ( \sum_{A,B,C} ( \frac{\mu_{gA}\mu_{AB}\mu_{BC}\mu_{Cg}}
{(\omega_A - \omega_{\sigma})(\omega_B - \omega_1 -
 \omega_2)(\omega_C - \omega_1)}+
\frac{\mu_{gA}\mu_{AB}\mu_{BC}\mu_{Cg}}
{(\omega_A^* + \omega_3)(\omega_B - \omega_1 -
 \omega_2)(\omega_C - \omega_1)}+ 
\nonumber \\
\frac{\mu_{gA}\mu_{AB}\mu_{BC}\mu_{Cg}}
{(\omega_A^* + \omega_1)(\omega_B^* + \omega_1 +
 \omega_2)(\omega_C - \omega_3)}+
\frac{\mu_{gA}\mu_{AB}\mu_{BC}\mu_{Cg}}
{(\omega_A^* + \omega_1)(\omega_B^* + \omega_1 + 
\omega_2)(\omega_C^* + \omega_{\sigma})})-
\nonumber \\
\sum_{A,C} ( \frac{\mu_{gA}\mu_{Ag}\mu_{gC}\mu_{Cg}}
{(\omega_A - \omega_{\sigma})(\omega_A - \omega_3)(\omega_C 
- \omega_1)}+ 
\frac{\mu_{gA}\mu_{Ag}\mu_{gC}\mu_{Cg}}
{(\omega_A - \omega_3)(\omega_C^* + \omega_2)(\omega_C - \omega_1)}+
\nonumber \\
\frac{\mu_{gA}\mu_{Ag}\mu_{gC}\mu_{Cg}}
{(\omega_A^* + \omega_{\sigma})(\omega_A^* + \omega_3)(\omega_C^* + 
\omega_1)}+ 
\frac{\mu_{gA}\mu_{Ag}\mu_{gC}\mu_{Cg}}
{(\omega_A^* + \omega_3)(\omega_C - \omega_2)(\omega_C^* + \omega_1)})),
\eea
where $\mu_{ij}$ is the dipole matrix element for the transition between 
the states $i$ and $j$, and $K(-\omega_\sigma;\omega_1,\omega_2,\omega_3)$ 
is a numerical constant which depends on the values of $\omega_{\sigma}$, 
$\omega_1$, $\omega_2$, $\omega_3$ \cite{orr71}.
In particular, for the EA coefficient 
($\omega_{\sigma} = - \omega_1 = - \omega, \omega_2 = \omega_3 = 0$), 
$K = 3$ and for the THG coefficient 
($\omega_{\sigma} = -3 \omega, \omega_1 = \omega_2 = \omega_3 = \omega$), 
$K = \frac{1}{4}$. 
$I_{1,2,3}$ denotes the average of all terms generated by permuting
$\omega_1$, $\omega_2$ and $\omega_3$.
A finite lifetime of the levels A, B, C is 
taken into account in order to calculate $\gamma_{xxxx}$ at the 
resonance points properly.

The sum in equation (11) is over all states. However, due to the 
fact that the ground state belongs to the $^1A_g^+$ symmetry sector, 
the dipole matrix elements are non-zero only for the transitions 
between $^1A_g^+$  and $^1B^-_{1u}$. Thus, the states A and C in (11)
are of $^1B_{1u}^-$ symmetry, while the state B, as well as the ground
state, are of $^1A_g^+$ symmetry.

Electroabsorption spectroscopy is used as a tool for investigating 
the low-lying electronic excitations in polymers (see, e.g., 
\cite{sebastian81-1,sebastian81-2,guo93-2,shimoi97}).
 In \cite{guo93} it was shown that 
EA spectrum includes features of two types: First, those similar to the 
derivative of the first-order optical absorption; they are located 
at the energies of the $^1\bu$ levels. A second type are the peaks 
at the energies of the $^1\ag$ levels, arising from the spatial 
symmetry breaking in the applied electric field.
Recent experimental studies of PPV and PPP EA spectra 
\cite{martin97,liess97,lane97} found both types of features in these systems.

Using the transition matrix elements we calculate the EA spectrum of 
PPP oligomers using the 10 lowest eigenstates.
The results for the 15 site oligomer are shown in Fig.\ 7. 
We note that transitions involving states of $B_{2u}$ and
$B_{3g}$ symmetry are not included in this figure.
The lowest-energy feature is a derivative-like feature arising from 
the $^1\bu$ level, showing a red shift. Next, the maximum corresponding 
to the $2^1\ag$ state is situated at about 4.3 eV. The second 
derivative-like feature almost coinciding with the 
maximum can be related to $n^1\bu$ state. Finally, the high energy
feature  at 
$4.8$ eV corresponds to the $k^1\ag$ state. 
The peaks corresponding to the $2^1\ag$ 
and $n^1\bu$ states are also identifiable
in the EA spectra of Lane {\em et al.} \cite{lane97} at approximately 
$4.2$ eV, just below the onset of the $1^1B_{2u}$ 
transitions. The high energy $k^1\ag$
feature is not identifiable in their data, owing to the $^1B_{2u}$
transitions.
For a comparison, results of the calculation with the five essential states
only are also shown. Evidently, these states describe
most of the EA spectrum.

Finally, we show the third harmonic generation (THG) spectrum of 
PPP oligomers \cite{referee}. This spectrum is related to the 
$\chi^{(3)}(-3\omega,\omega,\omega,\omega)$. Mazumdar $et$ $al.$ used 
the THG spectrum in their discussion of the essential states 
\cite{mazumdar94}. They argued that there are three main peaks in the 
THG spectra at E($1^1\bu$)/3, E($n^1\bu$)/3 and E($m^1\ag$)/2, 
where the
$m^1\ag$ is the $^1\ag$ state strongly coupled to the $1^1\bu$ 
exciton. Our calculations (shown in Fig.\ 8  for a 15 site oligomer) confirm
the existence of these peaks (those at 1.15, 1.5, and 2.1 eV, 
respectively), the $m^1\ag$ state being the $2^1\ag$ in our model. 
The peak at E($1^1\bu$)/3 is the highest in the spectrum.
However, the calculation also reveals an additional peak corresponding to 
the
$k^1\ag$ state at $2.4$ eV, as well as other peaks corresponding to
$^1\ag$ states. The feature at $3.4$ eV corresponds to a one-photon 
transition to the $1^1\bu$ state.
The essential states calculation shows features corresponding to
the essentials states but
fails to describe the total THG spectrum (employing 10 states)
as well as it does for the EA spectrum.

\section{Conclusions}

This paper has been
devoted to clarifying the nature of the low-lying 
$\ag$ and $\bu$ electronic excitations 
of PPP. A two state molecular-orbital
(2-MO) model, whose parameters are derived by 
fitting its predictions to exact Pariser-Parr-Pople model calculations 
of benzene and biphenyl, was used.
By solving this model using the DMRG method, 
it was shown that the lowest excited
state is a $^1\bu$ (`s'-wave) exciton, which is situated in the gap 
between the ground state and the band states. 
As the oligomer size increases, the
number of excitonic levels also increases, thus creating 
an excitonic band in the limit $N \rightarrow \infty$. 
The calculated
size of these excitons indicates that they are tightly bound, 
extending over only a few repeat units.
The lowest
band state is found to be a covalent 
$2^1\ag$ state, whose energy almost
coincides for all the oligomers studied with the charge gap $E_G$. 
Lying just above the $2^1\ag$ state is a band $^1\bu$ 
state (the $n^1\bu$ state).
The size of the
band states scales linearly with oligomer size. 
The binding energy of the $1^1\bu$ exciton is thus 
determined rigorously as $0.74$ eV.
The calculated values of the $1^1\bu$ and $2^1\ag$ energies of
$3.3$ and $4.0$ eV, respectively, in the long chain limit, are in good
agreement with the experimental results of $3.6$ and $4.2$ eV,
respectively.

The dipole matrix elements and oscillator strengths 
for the transitions between the lowest $^1\ag$ and $^1\bu$ states were
 calculated and 
the NLO properties of PPP, such as electroabsorption and 
third harmonic generation, were investigated. A comparison of 
the EA spectrum with the experimental data
\cite{lane97} showed that the main features of the experimental 
spectrum are well described in the 2-MO Hamiltonian. 
In particular, the derivative-like $1^1\bu$ feature and the 
peaks corresponding
to the band threshold $2^1\ag$ and $n^1\bu$ states are reproduced. 

Only five states account for most of the calculated electro-absorption. 
These are the $^1\ag$, $^1\bu$, $2^1\ag$, $n^1\bu$ and another band
$^1\ag$ state, the $k^1\ag$,
thus confirming the essential states model. An analysis of the 
particle excitation weight of these states indicates that  they are 
predominately single particle in character.

In conclusion, the parameterised 2-MO model presented here gives a
 quantitative
description of the low lying excitations and the NLO properties of
 the phenyl based
semiconductors. The parameterisation was performed to achieve
a good description of the
states within the $A_g$ and $B_{1u}$ symmetry sectors.  Equally,
however, by fitting to the relevant states of benzene and biphenyl,
a parameterisation could be achieved for states in 
the $B_{3g}$ and $B_{2u}$ symmetry sectors.

\section*{Acknowledgements}

The authors are grateful for valuable discussions with 
Professor D. Bradley, Professor G.\ Gehring,
 Dr.\ S.\ Martin and Dr.\ P.\ Lane 
from the University of Sheffield and with Dr.\ Y.\ Shimoi 
from the Electrotechnical Laboratory at Tsukuba, Japan.
One of the authors (M.\ Yu.\ L.) is supported by the 
EPSRC (U.\ K.) (grant ref. GR/K86343).
R.\ J.\ B.\ acknowledges the support of the Australian Research Council. 
The calculations were performed on the DEC8400 at the Rutherford 
Appleton Laboratory and on the SGI Power Challenge facility at the 
New South Wales Centre for Parallel Computing.

{\parindent=0pt
{\small{$^*$On leave from Institute of Inorganic Chemistry,
630090 Novosibirsk, Russia}}}

\vfill
\eject

\appendix

\section{Particle-hole Symmetry in the Molecular Orbital Basis} 

In this Appendix we derive the particle-hole inversion operator, 
$\hat{J}$, for the molecular orbital (MO) 
representation, and discuss the particle-hole-symmetry-adapted 
wavefunctions for a chain. In the atomic orbital (AO) representation 
the action of the particle-hole
inversion operator on the system can be cast as follows: 
\beq
J c^{\dagger}_{i\downarrow} = (-1)^i c_{i\uparrow},
\eeq
\beq
J c^{\dagger}_{i\uparrow} = (-1)^{i+1} c_{i\downarrow}.
\eeq
Using the following relations between the MO and AO creation operators:
\beq
a^{\dagger}_{1\sigma} =
 \frac{1}{\sqrt{12}} (2c^{\dagger}_{1\sigma}+
c^{\dagger}_{2\sigma}-c^{\dagger}_{3\sigma}-
2c^{\dagger}_{4\sigma}-c^{\dagger}_{5\sigma}+
c^{\dagger}_{6\sigma}), 
\eeq
\beq
a^{\dagger}_{2\sigma} = \frac{1}{\sqrt{12}}
 (2c^{\dagger}_{1\sigma}-c^{\dagger}_{2\sigma}-c^{\dagger}_{3\sigma}+
2c^{\dagger}_{4\sigma}-c^{\dagger}_{5\sigma}-
c^{\dagger}_{6\sigma}), 
\eeq
we deduce the following formulae for the action of the particle-hole 
inversion on the MOs as:
\beq
J a^{\dagger}_{1\uparrow} = a_{2\downarrow}, 
J a^{\dagger}_{1\downarrow} = - a_{2\uparrow}, 
\eeq
\beq
J a^{\dagger}_{2\uparrow} = a_{1\downarrow}, 
J a^{\dagger}_{2\downarrow} = - a_{1\uparrow}. 
\eeq

Now let us consider the action of the particle-hole operator within a 
single exciton basis. Yaron and Silbey have considered the action of 
$\hat{J}$ on a single exciton basis using periodic boundary conditions
\cite{yaron92}. 
Here, we develop these ideas using the real space MO representation. 

Let the ground state be represented by
\beq
\diracm{GS} = \prod_{i=1}^N a_{i1\uparrow}^{\dagger} 
a_{i1\downarrow}^{\dagger}\diracm{0}.
\eeq
Then we may create a singlet excitation of spatial extent 
$\delta= |i-j|$ and localised around the repeat unit $n=(i+j)/2$, as follows: 
\beq
\diracm{\psi_{n-\delta/2}^{n+\delta/2} } =S_{ij}^{\dagger}\diracm{GS},
\eeq
where
\beq
S_{ij}^{\dagger} = \frac{1}{\sqrt{2}} (a_{i2\uparrow}^{\dagger} a_{j1\uparrow}
 + a_{i2\downarrow}^{\dagger} a_{j1\downarrow}).
\eeq
It is straightforward to demonstrate that \beq
J
\diracm{\psi_{n-\delta/2}^{n+\delta/2} }= -S_{ji}^{\dagger}\diracm{GS}=-
\diracm{\psi_{n+\delta/2}^{n-\delta/2} }. \eeq
Thus, for an exciton localised about repeat unit $n$ we may construct a 
state with definite particle-hole symmetry as, 
\beq
\diracm{\phi_n^{\mp}}
= \sum_{\delta} \left( C_{\delta}^n
\diracm{\psi_{n-\delta/2}^{n+\delta/2}} \pm C_{\delta}^n
\diracm{\psi_{n+\delta/2}^{n-\delta/2}}
\right),
\eeq
where the symmetric (`s'-wave) combination is negative under $\hat{J}$, 
while the anti-symmetric (`p'-wave) combination is positive under $\hat{J}$. 
It is $C_{\delta}^n = C_{ij}$ which is essentially measured by eqn (5).

A state of definite $\hat{C}_2$ symmetry may then be constructed as follows: 
\beq
\diracm{\psi_j^{\mp}}=
\frac{1} {\sqrt{N+1}} \sum_n
\diracm{\phi_n^{\mp}}
\sin\left( \frac{n\pi j} {N+1} \right).
\eeq

\vfill
\eject

\newpage
\begin{center}
{\bf Figure Captions}
\end{center}
\begin{itemize}
\item[Figure 1 :] The structure of the poly($para$-phenylene) chain. 

\item[Figure 2 :]
Schematic representation of the superblocks used in the first two DMRG 
steps, showing the form of the system (S) and environment (E) blocks.

\item[Figure 3 :] Energies of the lowest excited $^1\ag$ and $^1\bu$ 
states relative to the ground state energy, and the charge gap $E_G$ 
(dashed line) as functions of the oligomer size, $N$. $1^1\bu$ 
(solid squares),
 $2^1\bu$ (solid diamonds), $3^1\bu$ 
(solid triangles), $4^1\bu$ (oblique crosses), $n^1\bu$ (empty diamonds) and
$2^1\ag$ (empty squares). 

\item[Figure 4 :] Mean electron-hole distance for singly-excited 
low-energy states as a function of oligomer size, $N$.
$1^1\bu$ (solid squares), $2^1\bu$ (solid diamonds), $3^1\bu$ 
(solid triangles), $n^1\bu$ (empty diamonds) and $2^1\ag$ 
(empty squares). 
Results in the absence of electron-electron interactions are shown in
dashed lines: $1^1\bu$ (solid squares) and $2^1\ag$ (empty squares). Also
shown are dashed lines without symbols corresponding to 
$N/2$ and $N/4$. 

\item[Figure 5 :] Essential states for the non-linear properties 
of the PPP and the most important one-photon transitions between them. 
The oscillator strengths are shown for a 15 site oligomer.

\item[Figure 6 :] Calculated first order optical absorption
spectrum from the $1^1\ag$ state to the
low-lying $^1\bu$ states for a 15 site oligomer.

\item[Figure 7 :] Calculated electroabsorption spectrum for a 15 site
 oligomer
within the $^1\ag$ and $^1\bu$ symmetry sectors.
 Solid diamonds: full calculation, empty diamonds: 
essential states calculation (see text).

\item[Figure 8 :] Calculated third harmonic generation spectrum for a 15 site
 oligomer
within the $^1\ag$ and $^1\bu$ symmetry sectors.
Solid diamonds: full calculation, empty diamonds: 
essential states calculation (see text). 

\end{itemize}

\vfill
\eject

\begin{table}

\centering

\begin{tabular}{|c|c|}
\hline
 Parameter & Value (eV) \\
\hline
$\Delta = \epsilon_2  - \epsilon_1$ & 5.26 \\
\hline
$t_{11}=t_{12}=-t_{21}=-t_{22}$ & 0.895 \\
\hline
$X=P$ & 0.89 \\
\hline
$ U = 2V$ & 3.67 \\
\hline 
\end{tabular}

\caption{ The values of the parameters used in the 2-MO model, eqn. (3).
}

\label{parameters}
\end{table}

\begin{table}

\centering

\begin{tabular}{|c|c|c|c|}
\hline Property & DMRG I & DMRG II & Exact \\ \hline $E_{GS}$ & 
-26.821231 & -26.821232 & -26.821232 \\ \hline $E_{1B_{1u}}$ &
 -23.239216 & -23.239302 & -23.239318 \\ \hline $E_{2A_g}$ 
& -22.254639 & -22.254786 & -22.254843 \\ \hline $E_{1B_{1u}} 
- E_{GS}$ & 3.582015 & 3.581930 & 3.581914 \\ \hline $E_{2A_g} 
- E_{GS}$ & 4.566592 & 4.566446 & 4.566389 \\ \hline R.m.s.($1B_{1u}$)
 & 1.400599 & 1.398925 & 1.396940 \\ \hline R.m.s.($2A_g$) & 2.660390 
& 2.652987 & 2.653227 \\ \hline
\end{tabular}

\caption{Energies of the ground and lowest excited states, as well as 
exciton sizes of the lowest excited states of a $N = 7$ chain for exact 
and DMRG calculations (DMRG I: 65,000 states, DMRG II: 
130,000 states).} 

\label{comparison}
\end{table}


\begin{table}[h]

\centering

\begin{tabular}{|c|c|c|c|c|}
\hline N & $E_0$ (EXACT) & $E_0$ (DMRG) & $E_1$ (EXACT) & $E_1$ (DMRG) \\
\hline 3 & $-16.9946597$ & $-16.9946597$ & $-13.9110457$ & $- 13.9110457$ \\
\hline 5 & $-28.7708529$ & $-28.7708529$ & $-26.3070239$ & $- 26.3070239$ \\
\hline 7 & $-40.5492152$ & $-40.5492152$ & $-38.3620419$ & $- 38.3620023$ \\
\hline 9 & $-52.3277963$ & $-52.3277961$ & $-50.2911319$ & $- 50.2910587$ \\
\hline 11 & $-64.1064055$ & $-64.1064050$ & $-62.1614635$ & $- 62.1613673$ \\
\hline 13 & $-75.8850187$ & $-75.8850178$ & $-74.0003255$ & $- 74.0002074$ \\
\hline 15 & $-87.6636326$ & $-87.6636311$ & $-85.8207064$ & $- 85.8205619$ \\
\hline 17 & $-99.4422467$ & $-99.4422442$ & $-97.6294862$ & $- 97.6293092$ \\
\hline 19 & $-111.220861$ & $-111.220857$ & $-109.430604$ & $- 109.430389$ \\
\hline
\end{tabular}

\caption{Ground and first excited state energies 
 in the non-interacting limit for various 
oligomer lengths $N$ as calculated exactly and using the DMRG method.} 
\label{dmrgtest1}
\end{table}

\begin{table}[h]

\centering

\begin{tabular}{|c|c|c|c|c|c|c|}
\hline N & $m=70$ & $m=100$ & $m=130$ & $m=160$ & $m=175$ & $m=210$ \\
\hline 7 & 3.583 & 3.582 & 3.582 & 3.581 & 3.581 & 3.581 
\\ \hline 9 & 3.480 & 3.477 & 3.477 & 3.477 & 3.477 & 3.477
 \\ \hline 11 & 3.424 & 3.420 & 3.420 & 3.420 & 3.420 & 3.420 
\\ \hline 13 & 3.392 & 3.387 & 3.386 & 3.386 & 3.386 & 3.386 
\\ \hline 15 & 3.373 & 3.365 & 3.365 & 3.365 & 3.364 & 3.364 
\\ \hline 17 & 3.361 & 3.351 & 3.351 & 3.351 & 3.350 & 3.350 
\\ \hline 19 & 3.353 & 3.341 & 3.341 & 3.341 & 3.340 & 3.340 
\\ \hline 21 & 3.348 & 3.334 & 3.334 & 3.334 & 3.333 & 3.333 
\\ \hline
\end{tabular}

\caption{Values in eV of the exciton gap, the difference between 
the $(1^{1}B_{1u}^{- })$ and $(1^{1}A_{g}^{+})$ energies, as the 
truncation parameter $m$ is increased, for a range of oligomer 
lengths $N$.} 
\label{dmrgtest2}
\end{table}

\begin{table}[h]

\centering

\begin{tabular}{|c|c|c|c|c|c|c|c|c|c|}
\hline $N$ & $1^1B^-_{1u}$ & $2^1A^+_g$ & $1^3B^+_{1u}$ 
& Experimental Optical Gap \\
\hline 2 & 4.80 & 6.17 & 3.55 & 4.80(a)\\
 \hline 3 & 4.28 & 5.60 & 3.17 & 4.5(b) \\
 \hline 4 & 3.98 & 5.18 & 2.97 & --- \\
\hline 5 & 3.79 & 4.89 & 2.85 & --- \\
\hline 6 & 3.67 & 4.70 & 2.78 & 3.9(b) \\
 \hline 7 & 3.58 & 4.57 & 2.73 & --- \\
\hline 11 & 3.42 & 4.29 & 2.65 & --- \\
\hline 13 & 3.39 & 4.22 & 2.63 & --- \\
\hline 15 & 3.36 & 4.18 & 2.62 & --- \\
\hline 17 & 3.35 & 4.14 & 2.61 & --- \\
\hline 19 & 3.34 & 4.12 & 2.60 & --- \\
\hline 21 & 3.33 & 4.11 & 2.60 & --- \\
\hline $\infty$ & 3.30 & 4.04 & 2.60 & 3.43(b), 3.3(c), 3.5(d), 3.8(e) \\
\hline
\end{tabular}

\caption{Calculated vertical transition energies in eV for oligophenylenes of
various lengths $N$. Note that
the
$2^1A^+_g$ exciton is the {\em lowest covalent} $A_g$ singlet excited state.
Experimental results from biphenyl crystals (a) \citex{crystal} and crystalline
films (b) \citex{shacklette}, (c) \citex{tieke}, (d) \citex{ambroschdraxl95}, 
(e) \citex{lane97}.}

\label{ppp2}
\end{table}

\begin{table}

\centering

\begin{tabular}{|c|c|c|c|}
\hline $1^1\bu$&$n^1\bu$ $(=4^1\bu$)&$m^1\ag$ $(=2^1\ag)$&
$k^1\ag$ $(=4^1\ag)$ \\
\hline 0.996 & 0.878 & 0.794 & 0.797 \\
\hline

\end{tabular}

\caption{The single particle excitation weight, $W_1$ (7), of the essential
states for a 15 site oligomer.}
\label{weights}
\end{table}

\begin{table}

\centering

\begin{tabular}{|c|c|c|c|c|c|}
\hline $N$ & $1^1\ag \rightarrow 1^1\bu$ & $1^1\bu \rightarrow 2^1\ag$ &
 $2^1\ag \rightarrow n^1\bu$&$1^1\ag \rightarrow n^1\bu$ & $n^1\bu
 \rightarrow k^1\ag$ \\
\hline 3 & 21.330 (4.28) &
 7.091 (1.32) & 26.391 (2.23) & 0.025 (7.83) & 21.202 (1.43) \\ 
\hline 5 & 35.858 (3.79) &
 12.963 (1.10) & 48.603 (1.48) & 0.314 (6.37) & 30.152 (1.04) \\ 
\hline 7 & 49.237 (3.58) &
 21.375 (0.99) & 66.291 (1.01) & 0.885 (5.58) & 40.231 (0.39) \\ 
\hline 9 & 62.077 (3.48) &
 28.833 (0.92) & 79.324 (0.72) & 1.330 (5.12) & 77.925 (0.33) \\ 
\hline 11 & 74.603 (3.42) &
 34.015 (0.87) & 87.975 (0.54) & 1.664 (4.83) & 107.969 (0.29) \\ 
\hline 13 & 87.109 (3.39) &
 37.276 (0.84) & 93.555 (0.41) & 1.760 (4.64) & 124.494 (0.24) \\ 
\hline 15 & 99.662 (3.37) &
 39.144 (0.82) & 97.084 (0.33) & 1.796 (4.51) & 133.758 (0.21) \\ 
\hline 17 & 112.411 (3.35) &
 40.027 (0.80) & 99.091 (0.27) & 1.727 (4.42) & 140.365 (0.18) \\ 
\hline 19 & 125.153 (3.34) &
 40.247 (0.79) & 100.176 (0.22) & 1.667 (4.36) & 144.788 (0.15) \\ 
\hline 21 & 138.013 (3.34) &
 40.142 (0.78) & 100.612 (0.19) & 1.590 (4.31) & 146.118 (0.13) \\ 
\hline
\end{tabular}

\caption{Calculated oscillator strengths for selected transitions in 
oligophenylenes of various sizes, $N$. (The corresponding energy 
differences in eV are shown in brackets.)}
\label{oscstr}
\end{table}


\begin{thebibliography}{40}

\bibitem{nature90} J.\ H.\ Burroughes, D.\ D.\ C.\ Bradley, A.\ R.\ Brown,
R.\ N.\ Marks, K.\ Mackay, R.\ H.\ Friend, P.\ L.\ Burns 
and A.\ B.\ Holmes,
Nature {\bf 347}, 539 (1990).

\bibitem{grem92} G.\ Grem, G.\ Leditzky, B.\ Ullrich and G.\ Leising, 
Adv. Mater. {\bf 4}, 36 (1992).

\bibitem{ambroschdraxl95} C.\ Ambrosch-Draxl, J.\ A.\ Majewski, 
P.\ Vogl and G.\ Leising, Phys. Rev. B {\bf 51}, 9668 (1995). 

\bibitem{lane97} P.\ A.\ Lane, M.\ Liess, Z.\ V.\ Vardeny, M.\ Hamaguchi,
M.\ Ozaki and K.\ Yoshino, Synth. Met. {\bf 84} 641 (1997). 

\bibitem{barford97a} W.\ Barford and R.\ J.\ Bursill, Chem. Phys. Lett. 
{\bf 268}, 535 (1997).

\bibitem{barford97b} W.\ Barford and R.\ J.\ Bursill, Synth. Met.
{\bf 89 }, 155  (1997)

\bibitem{bb3}
R.\ J.\ Bursill, C.\ Castleton and W.\ Barford {\em Optimal
Parameterisation of the Pariser-Parr-Pople
Model for Benzene and Biphenyl}, submitted to Chem.\ Phys.\ Lett.\ (1997).

\bibitem{bredas2}
J.\ L.\ Br\'{e}das, J.\ Chem.\ Phys.\ 82 (1985) 3808. 

\bibitem{champagne96} B.\ Champagne, D.\ H.\ Mosley, J.\ G.\ Friiat and 
J.\-M.\ Andre,
Phys.\ Rev.\ B 54 (1996) 2381.

\bibitem{rice}
Yu.\ N.\ Gartstein, M.\ J.\ Rice and E.\ M.\ Conwell,
 Phys.\ Rev.\ B 52 (1995) 1683.

\bibitem{shimoi96} Y.\ Shimoi and S.\ Abe, Synth. Met. 78, 219 (1996).

\bibitem{shimoi97} Y.\ Shimoi and S.\ Abe, Synth. Met. {\bf 91}, 363 (1997).

\bibitem{harigaya97} K.\ Harigaya, J. Phys.: Condens. Matt.
 {\bf 9}, 5989 (1997).

\bibitem{orr71} B.\ J.\ Orr and J.\ F.\ Ward, Mol. Phys. {\bf 20}, 513 (1971).

\bibitem{dixit91} S.\ N.\ Dixit, D.\ Guo and S.\ Mazumdar, 
Phys. Rev. B {\bf 43}, 6781 (1991).

\bibitem{guo93} D.\ Guo, PhD thesis, University of Arizona, 1993. 

\bibitem{mazumdar94} S.\ Mazumdar and F.\ Guo, J. Chem. Phys. {\bf 100}, 
1665 (1994).

\bibitem{yaron96} D.\ Yaron, Phys. Rev. B {\bf 54}, 4609 (1996). 


\bibitem{soos92} Z.\ G.\ Soos, S.\ Etemad, D.\ S.\ Galvao 
and S.\ Ramasesha,
Chem. Phys. Lett. {\bf194}, 341 (1992).

\bibitem{chandross97} M.\ Chandross, Y.\ Shimoi and S.\ Mazumdar, 
Synth. Met. {\bf 85}, 1001 (1997), Chem.\ Phys.\ Lett.\ 280, 85 (1997).

\bibitem{bursill97} R.\ J.\ Bursill, W.\ Barford and H.\ Daly
{\em Molecular Orbital Models of Benzene, Biphenyl and the
Oligophenylenes}, submitted to Chemical Physics (1997).

\bibitem{biphenylnote} Note that the $2^1\ag$ state is the first
excited {\em covalent} $A_g$ state of biphenyl.  There is a
$B_{3g}^+$ state (at ca.\ $4.11$ eV) and an {\em ionic} $A_g$
state (at ca.\ $5$ eV) below it.

\bibitem{white} S.\ R.\ White, Phys.\ Rev.\ Lett.\ {\bf 69}, 2863 (1992); 
Phys.\ Rev.\ B {\bf 48}, 10354 (1993). 

\bibitem{gehring}
G.\ A.\ Gehring, R.\ J.\ Bursill and T.\ Xiang, Acta Physica Polonica 
91 (1997) 105.

\bibitem{crystal}
T.\ G.\ McLaughlin and L.\ B.\ Clark, Chem.\ Phys.\ 31 (1978) 11.

\bibitem{shacklette}
L.\ W.\ Shacklette, H.\ Eckhardt, R.\ R.\ Chance, G.\ G.\ Miller, 
D.\ M.\ Ivory and R.\ H.\ 
Baughman, J.\ Chem.\ Phys.\ 73 (1980) 4098.

\bibitem{tieke}
B.\ Tieke, C.\ Bubek and G.\ Lieser, Makroml.\ Chem.\, Rapid Comm.\
 3 (1982) 261.

\bibitem{yaron92} D.\ Yaron and R.\ Silbey, 
Phys. Rev. B {\bf 45}, 11,655 (1992).

\bibitem{liess97} M.\ Liess, S.\ Jeglinski, Z.\ V.\ Vardeny, 
M.\ Ozaki, K.\ Yoshino, Y.\ Ding and T.\ Barton, Phys. Rev. B
{\bf 56}, 15712 (1997). 

\bibitem{neher} D.\ Neher, W.\ E.\ Torruellas, K.\ B.\ Rochford,
M.\ B.\ Marques, R.\ Zanoni, G.\ Assanto and G.\ I.\ Stegeman,
Synth. Met. {\bf 49-50}, 21 (1992).

\bibitem{sebastian81-1} L.\ Sebastian and G.\ Weiser, Chem. Phys. {\bf 61},
125 (1981).

\bibitem{sebastian81-2} L.\ Sebastian and G.\ Weiser, Phys. Rev. Lett. 
{\bf 46}, 1156 (1981).

\bibitem{guo93-2} D.\ Guo, S.\ Mazumdar, S.\ N.\ Dixit, F.\ Kajzar, 
F.\ Jarka, Y.\ Kawabe and N.\ Peyghambarian, Phys. Rev. B {\bf 48},
 1433 (1993).

\bibitem{martin97} S.\ J.\ Martin, H.\ Mellor, D.\ D.\ C.\ Bradley 
and P.\ L.\ Burn, Optical Materials {\bf 9}, 88 (1998).

\bibitem{referee} We would like to thank one of the referees for drawing our
attention to an error in the calculation of the THG spectrum 
in the initial version of the manuscript.


\end{thebibliography}
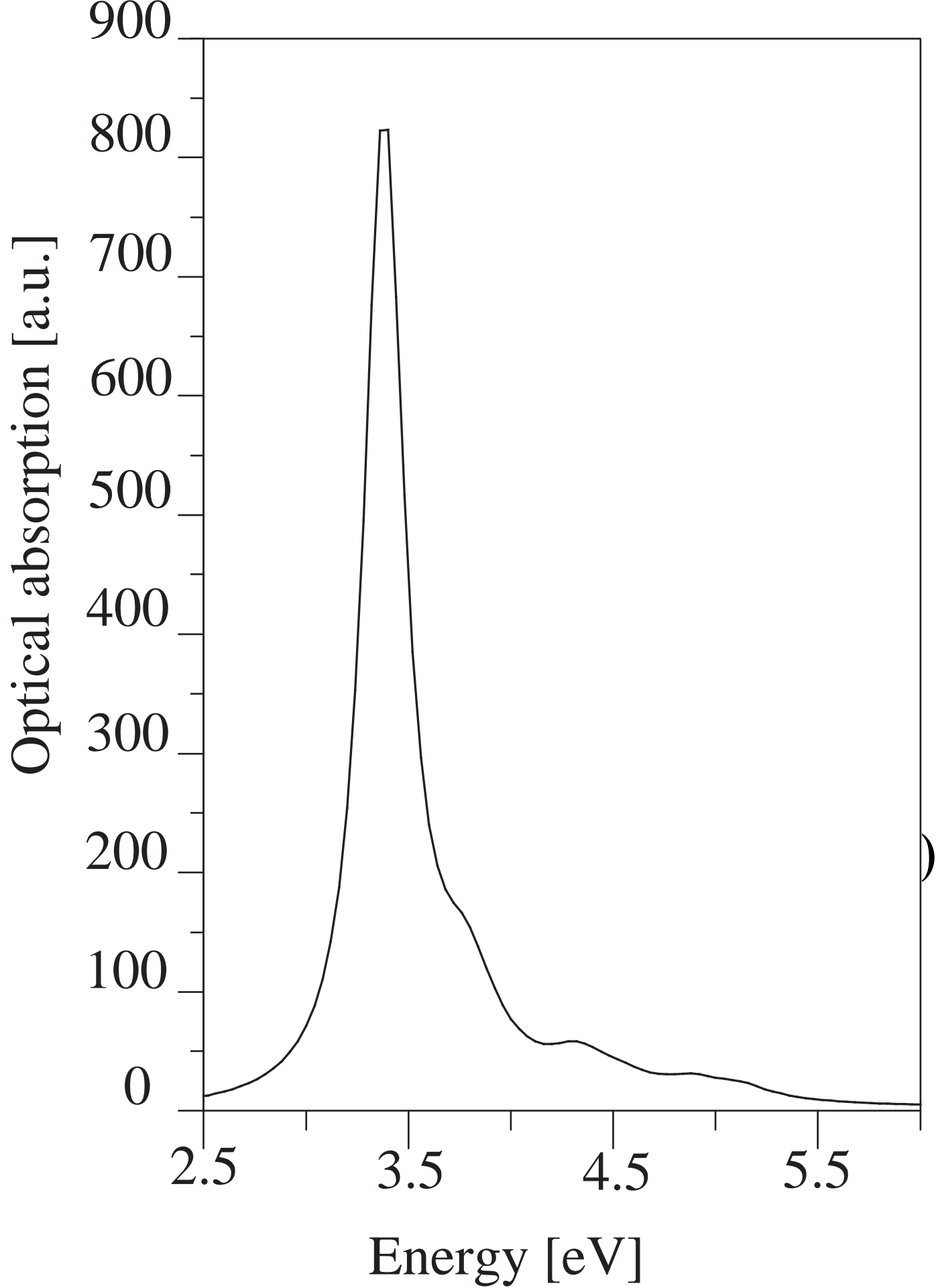
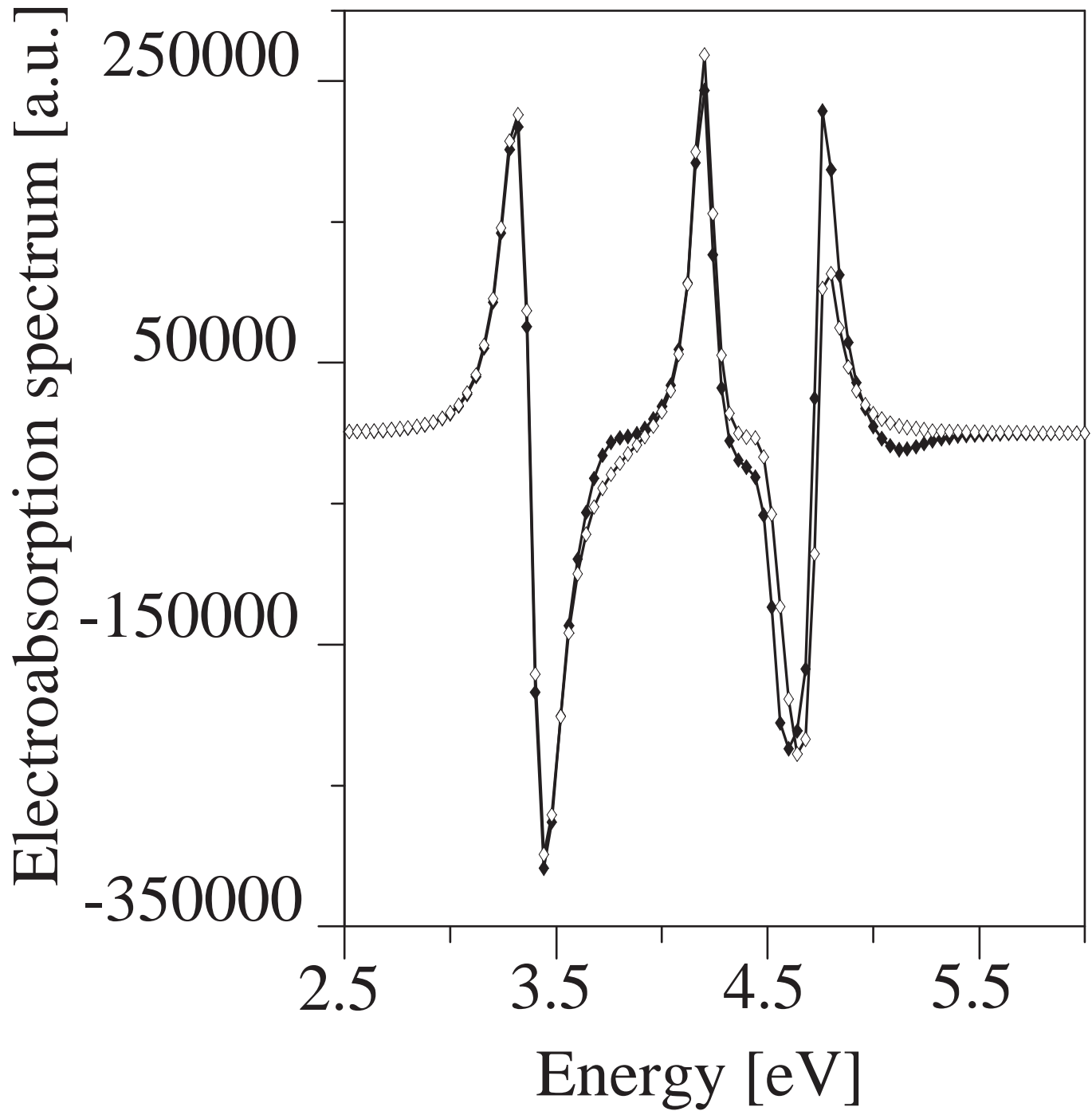
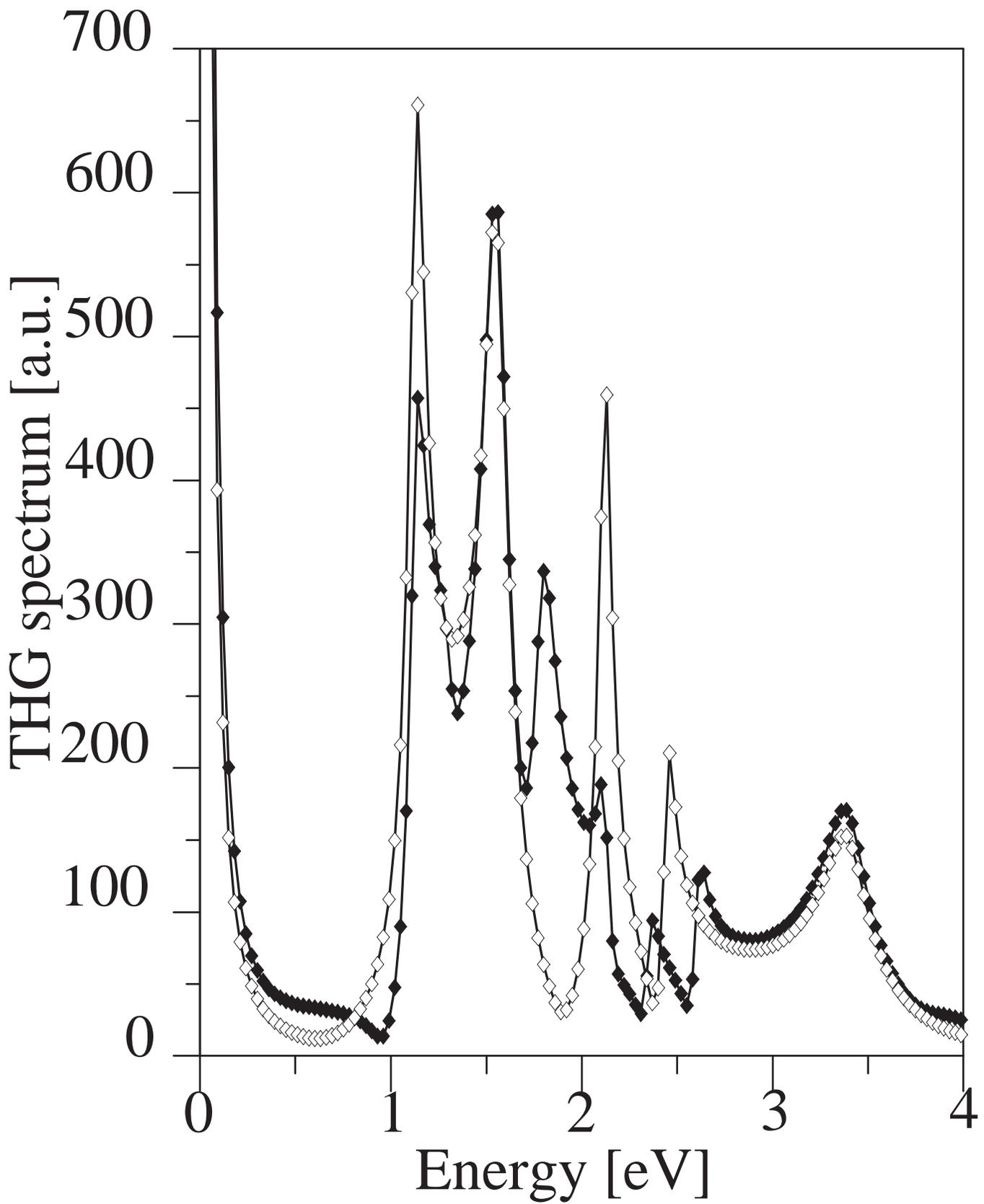
\end{document}